\documentclass[twocolumn,showpacs,amsmath,amssymb,superscriptaddress,aps,prc]{revtex4-1}
\usepackage[dvips]{graphicx}
\usepackage{epsfig}
\usepackage{amsmath}
\usepackage[dvips]{color}
\newcommand{\vis}[1]{\mbox{\boldmath $#1$}}
\newcommand{\cn}{\mbox{$^{20}{\rm C}+n$}}
\newcommand{\cnn}{\mbox{$^{20}{\rm C}+n+n$}}
\newcommand {\carba} {\mbox{$^{22}$C}}
\newcommand {\carbb} {\mbox{$^{21}$C}}
\newcommand {\kmax} {\mbox{$K_{\rm max}$}}

\begin{document}
\title{Coulomb breakup of $^{22}$C in a four-body model}
\author{E. C. Pinilla}
\affiliation{Universidad Nacional de Colombia, Sede Bogot\'a, Facultad de Ciencias, Departamento de F\'\i{}sica, Grupo de F\'\i{}sica Nuclear, Carrera 45 N\textsuperscript{o} 26-85, Edificio Uriel Guti\'errez, Bogot\'a D.C. C.P. 1101, Colombia.}
\author{P. Descouvemont}
\affiliation{Physique Nucl\'eaire Th\'eorique et Physique Math\'ematique, C.P. 229,\\
Universit\'e Libre de Bruxelles (ULB), B 1050 Brussels, Belgium}
\date{\today}
\begin{abstract}
Breakup cross sections are determined for the Borromean nucleus $^{22}$C by using a four-body eikonal model, including Coulomb corrections. 
Bound and continuum states are constructed within a \cnn\ three-body model in hyperspherical coordinates. We compute continuum states with the correct asymptotic behavior through the $R$-matrix method. 
For the $n+n$ potential, we use the Minnesota interaction. As there is no precise experimental 
information on $^{21}$C, we define different parameter sets for the $\cn$ potentials.
These parameter sets provide different scattering lengths, and resonance energies of an 
expected $3/2^+$ excited state. Then we analyze the $^{22}$C ground-state energy and rms radius,
as well as E1 strength distributions and breakup cross sections. The E1 strength distribution presents an enhancement at low energies. Its amplitude is associated with the low binding energy, rather than with a three-body resonance.
We show that the shape of the cross section at low energies is sensitive to the ground-state properties. 
In addition, we suggest the existence of a low-energy $2^+$ resonance, which should be observable in 
breakup experiments.
\end{abstract}
\maketitle
\section{Introduction}
One of the main characteristics of halo nuclei is their anomalously large radii, in comparison with their isotopic neighbors. They also 
present enhanced electric dipole distributions at low excitation energies, which seems to be an universal property. However, it is still debated if this property is related with a resonance behavior \cite{DPB12,PDB12} or if it is an effect coming from the weak binding of the ground state \cite{SLY03,NLV05}.

Among halo nuclei, Borromean nuclei are made of three-body structures, a core and two loosely bound nucleons. They present a weakly bound state only, and no pair core-nucleon or nucleon-nucleon is bound. Typical examples are $^{11}$Li=$^{9}$Li$+n+n$, $^6$He$=^{4}$He$+n+n$ and $^{14}$Be$=^{12}$Be$+n+n$. 

$^{22}$C is the heaviest Borromean nucleus known so far. Tanaka et al. \cite{TYS10} deduced a very large rms matter radius 
($r_{rms}=5.4\pm 0.9$ fm), and infer a two neutron separation energy, $S_{2n}=0.42\pm0.94$ MeV, using a simplified three-body model. A 
 recent mass measurement limits $S_{2n}$ to $S_{2n}<300$ keV \cite{GMO12}. Little information is known about the ground-state energy of $^{22}$C. 

In three-body nuclei, the understanding of two-body subsystems is crucial. 
$^{21}$C is known to be unbound with little experimental spectroscopic information available. Mosby {\sl et al.} \cite{MBB13} give a limit to the scattering length,  $|a_0|<2.8$ fm, through one proton removal from $^{22}$N. From this result and a zero-range renormalized three-body model \cite{YMF11,YMF12}, these authors provide  $S_{2n}<70$ keV. Even if accurate three-body models are currently available, 
the absence of well-established information on $^{21}$C limits three-body calculations of $^{22}$C. 

Three-body calculations of $^{22}$C have been performed in Refs. \cite{HS06,KT14,EVZ12,OMF13} assuming a \cnn\ structure for
 the ground state. In Ref.\ \cite{HS06}, $\cn$ deep potentials are constructed and they are determined in such a way that different energies of the single particle $0d_{5/2}$ state are provided. The Pauli principle is approximately taken into account considering that the bound states in the $0s_{1/2}$, $0p_{3/2}$, $0p_{1/2}$ and $0d_{5/2}$ orbits simulate the forbidden states. 
 In Ref.\ \cite{EVZ12}, $l$-independent $\cn$ potentials that do not support forbidden states are used to study, in a simple approach, the relation between the $r_{\text{rms}}$ radius and the ground state energy of $^{22}$C, with the E1 strength distribution. Different sets of potentials with $l$-dependent central parts are considered in Ref. \cite{KT14} to calculate reaction cross sections of $^{22}$C on $^{12}$C at 300 MeV/nucleon. The relation between the scattering length of the $1s_{1/2}$ state and the ground state energy of $^{22}$C is shown. However a three-body phenomenological force \cite{DTV98} is added to the Hamiltonian, which hides the direct link between the two-body scattering length and the three-body ground state energy.

Breakup experiments are typically performed at energies much higher than the Coulomb barrier, where eikonal models are suitable. They consist in high energy approximations that reduce the Schr\"odinger equation, a second order differential equation, to a first order one, which constitutes a strong simplification in four-body calculations. 
Assuming a Coulomb E1 dominated breakup process, the breakup of halo nuclei can be directly related with the E1 strength distribution through the equivalent photon method \cite{BB88}. This method simplifies the calculation of the breakup excitation function. However, the inclusion of contributions other than dipole could be important in analysing experimental data \cite{PDB12}.

A four-body eikonal calculation, including Coulomb corrections, has been applied to determine elastic and breakup cross sections 
of $^6$He \cite{BCD09} and $^{11}$Li \cite{PDB12} on $^{208}$Pb.  This model is more appropriate than the equivalent photon method, which 
is traditionally used for experimental \cite{NVS06} and theoretical \cite{KKM10,HS07} studies of Coulomb breakup. The present model is more accurate since:  $i)$ it involves three-body continuum wave functions with the correct asymptotic behavior; $ii)$ multipolarities different from dipole can be taken into account; $iii)$ Coulomb and nuclear effects, and their interference, are introduced consistently; $iv)$ E1 strength distributions and the breakup cross sections are computed separately.

The aim of the present work is to apply a four-body reaction model to study the Coulomb breakup of $^{22}$C. Bound and continuum states are defined
 in hyperspherical coordinates \cite{DDB03,DTB06}. Continuum \cnn\ three-body states are computed with the correct asymptotic behavior through the $R$-matrix method \cite{DTB06}. We calculate the breakup cross section for a $^{22}$C projectile impinging on $^{208}$Pb at 240 MeV/nucleon, an energy typical of the energies available at RIKEN. As there is still a significant experimental uncertainty on the
 binding energy  of $^{22}$C, we consider different conditions of the calculations, corresponding to various energies.
 
Since we do not have precise experimental information on $^{21}$C, we construct $\cn$ deep $l-$dependent potentials to study three-body properties of $^{22}$C, the ground-state energy and $r_{\text{rms}}$ radius.
The $\cn$ potentials are consistent with experimental information, i.e. the scattering length of a $1s_{1/2}$ virtual state and the energy of a possible $0d_{3/2}$ resonance \cite{N15} in $^{21}$C. 

The paper is organized as follows: Section \ref{theory} briefly describes the four-body eikonal model and the hyperspherical formalism to construct three-body wave functions. In Section \ref{GSproperties}, we study ground state properties when different $\cn$ potentials are chosen. In Section \ref{Breakup}, we determine electric dipole strength distributions and breakup cross sections.  Summary and conclusions are given in Section \ref{conclusions}.

\section{The three-body model}
\label{theory}
\subsection{The $^{22}$C nucleus in hyperspherical coordinates}
Before introducing the eikonal model, let us describe the three-body model of the projectile used to compute the bound and scattering states involved in the breakup cross sections. Here we just outline the three-body model. For details see for instance Refs. \cite{ZDF93,DDB03}.

The Hamiltonian for a three-body nucleus, consisting of three clusters with nucleon numbers $A_i$, is given by
\begin{equation}
H_{3b}=\sum\limits_{i=1}^{3}\frac{\vis{p^2_i}}{2m_NA_i}+\sum\limits_{i<j=1}^{3}V_{ij}(\vis{r_i}-\vis{r_j}),
\label{Ham3b}
\end{equation}
where $m_N$ is the nucleon mass, $\vis{r_i}$ and $\vis{p_i}$ are the space coordinate and momentum of nucleus $i$,  
and $V_{ij}$ an interaction between the nuclei $i$ and $j$. 
For a three-body nucleus made of a core and of two nucleons, we define the scaled Jacobi coordinates by 
\begin{gather}
\vis{x}=\frac{1}{\sqrt{2}}\left(\vis{r}_3-\vis{r}_2\right),\notag\\
\vis{y}=\sqrt{\frac{2A_1}{A_1+2}}\left(\vis{r}_1-\frac{\vis{r}_2+\vis{r}_3}{2}\right),
\label{Jacobi}
\end{gather}
$\vis{r_1}$ being the space coordinate of the core of mass number $A_1$, and $\vis{r_2}$ and $\vis{r_3}$ being the space nucleon coordinates. The set of coordinates (\ref{Jacobi}) corresponds to the so called \textquotedblleft T-basis". The \textquotedblleft Y-bases" are defined by cyclic permutations of the core and nucleon coordinates \cite{DDB03}. Transformations between the different bases can be performed through Raynal-Revai coefficients \cite{RR70}. 
 
The hyperspherical coordinates are defined from the scaled Jacobi coordinates by
\begin{eqnarray}
\rho^2=x^2+y^2,\qquad
\alpha=\arctan\frac{y}{x};\quad 0\leq\alpha \leq \frac{\pi}{2},
\end{eqnarray}
where $\rho$ is called the hyperradius and $\alpha$ the hyperangle.

A partial wave solution of the three-body Schr\"odinger equation associated with the Hamiltonian (\ref{Ham3b}), with total angular momentum $J$, projection $M$ and parity $\pi$ can be expanded in hyperspherical coordinates as
 \begin{eqnarray}
\Psi^{JM\pi}(\rho,\Omega_{5\rho})=\rho^{-5/2}\sum_{K=0}^{\infty}\sum_{\gamma} \chi^{J\pi}_{\gamma K}(\rho) {\cal Y}^{JM}_{\gamma K}(\Omega_{5\rho}).
\label{pwf}
\end{eqnarray}
In Eq.\ (\ref{pwf}), $\gamma$ stands for $\gamma=(l_x, l_y, L, S)$, ${\cal Y}^{JM}_{\gamma K}(\Omega_{5\rho})$ is an hyperspherical harmonics \cite{DDB03} with $\Omega_{5\rho}=(\Omega_x,\Omega_y,\alpha)$ and $\Omega_x$, $\Omega_y$ are the solid angles of the $\vis{x}$ and $\vis{y}$ scaled Jacobi coordinates, respectively. The function $\chi^{J\pi}_{\gamma K}(\rho)$ is called hyperradial wave function.

The angular momenta are coupled as
\begin{align}
&|l_x-l_y|  \leq L \leq l_x+l_y, \nonumber \\
&|S_1-S_2|\leq S\leq S_1+S_2,\nonumber \\
&|L-S|\leq J\leq L+S,
\end{align}
where $l_x$ and $l_y$ are the orbital quantum numbers associated with the scaled Jacobi coordinates $\vis{x}$ and $\vis{y}$, and 
$S_1=S_2=1/2$ are the intrinsic spins of the nucleons. 
The hypermomentum quantum number $K$ is defined as
\begin{equation}
K=2n+l_x+l_y,
\end{equation}
$n$ being a positive integer.
In practice the sum in Eq.\ (\ref{pwf}) is truncated up to a $K_{\text{max}}$ value and the parity $\pi=(-1)^K$ limits this sum to even or odd values.

Inserting expansion (\ref{pwf}) in the three-body Schr\"odinger equation provides the set of coupled differential equations
\begin{equation}
\left(T_K-E \right)\chi^{J\pi}_{\gamma K}(\rho)
+\sum_{K' \gamma'}V^{J\pi}_{\gamma K ,\gamma' K'}(\rho)\, \chi^{J\pi}_{\gamma' K'}(\rho)=0,
\label{cde}
\end{equation}
where the kinetic-energy operator is defined as
\begin{equation}
T_K=-\frac{\hbar^2}{2m_N}
\left[\dfrac{d^2}{d\rho^2}-\dfrac{(K+3/2)(K+5/2)}{\rho^2}\right]
\end{equation}
and $V^{J\pi}_{\gamma K ,\gamma' K'}(\rho)$ is a matrix element of the total potential $V_{12}+V_{13}+V_{23}$ between hyperspherical harmonics \cite{DDB03}.

The hyperradial bound-state wave functions are obtained variationally i.e. through the expansion 
\begin{equation}
\chi^{J\pi}_{\gamma K}(\rho)=\sum\limits_{i=1}^N c_{\gamma Ki}^{J\pi}\varphi_i(\rho).
\label{chivar}
\end{equation}
We use a Lagrange basis \cite{Ba15} as the set of $\varphi_i$ . This basis is made of orthonormal functions that vanish at all points of an associated mesh except at one. When the Hamiltonian matrix elements are computed at the Gauss approximation, one gets analytical matrix elements of the kinetic operator and diagonal matrix elements of the potential evaluated at the mesh points. Thus, the use of a Lagrange basis simplifies in great amount the numerical calculations since we do not need to perform integrals for the matrix elements.

Continuum states are defined as in Eq. (8) of Ref. \cite{BCD09}. At large distances, the nuclear potential is negligible. 
The hyperradial wave functions therefore behave as
\begin{align}
\chi^{J\pi}_{\gamma K(\gamma ' K')}(E,\rho) \underset{\rho \rightarrow \infty}\longrightarrow 
i^{K'+1}&(2\pi /\kappa )^{5/2}\bigl [H^-_{\gamma K}(\kappa \rho)\delta_{\gamma \gamma'}\delta_{KK'}\notag\\
&-U^{J\pi}_{\gamma K,\gamma ' K'}H^+_{\gamma K}(\kappa \rho) \bigr],
\label{asympchi}
\end{align}
where $H^\pm_{\gamma K}(x)$ are Hankel functions \cite{AS72}, $\kappa=\sqrt{2m_NE/\hbar^2}$ is the wave number,
 and $U^{J\pi}_{\gamma K,\gamma ' K'}$ is the three-body collision matrix. 
Indices $K' \gamma'$ define the entrance channel.
Here, $E>0$ is the excitation energy of the projectile defined from the three-body breakup threshold.

We use the three-body $R$-matrix method \cite{DTB06} to find the continuum states with the appropriate asymptotic behavior (\ref{asympchi}). This method consists in dividing the configuration space into two regions, the internal region, where the hyperradial wave function is expanded over basis (\ref{chivar}), and the external region, where the wave function is given by Eq.\  (\ref{asympchi}). From the matching of the wave functions at the boundary of the two regions one finds the collision matrix and the coefficients $c_{\gamma Ki}^{J\pi}$ that define the hyperradial wave function in the internal region.

\subsection{E1 strength distribution}
For a system made of a core and two halo neutrons, the electric dipole operator is defined as
\begin{equation}
\mathcal{M}^{E1}_{\mu}(\alpha,\rho)=eZ_1\left(\frac{2}{(2+A_1)A_1}\right)^{1/2}\rho\sin\alpha Y_1^\mu(\Omega_y),
\label{emop}
\end{equation}
with $Z_1$ the charge number of the core.

The distribution of transition probabilities from the bound state to the continuum through the dipole electric operator (\ref{emop})
 is given by
\begin{align}
\frac{dB(E1)}{dE}=&\frac{1}{2J_0+1}
\times\nonumber\\
&\sum\limits_{S\nu M_0\mu}\int d\vis{k_x}d\vis {k_y} \, \delta\left[E-\frac{\hbar^2}{2m_N}
(k_x^2+k_y^2)\right]\times\nonumber\\
&\left|\langle\Psi^{(-)}_{\vis {k_x},\vis{k_y},S\nu}(E,\vis{x},\vis{y})|\mathcal{M}^{(E1)}_\mu|\Psi^{J_0M_0\pi_0}(\vis{x},\vis{y})\rangle\right|^2,
\label{E1sexact}
\end{align}
where $\Psi^{J_0M_0\pi_0}(\vis{x},\vis{y})$ is the initial ground state defined as in Eq. (\ref{pwf}), with total angular momentum $J_0$, projection on the $z$ axis $M_0$ and parity $\pi_0$. The time-reversed continuum state is represented by $\Psi^{(-)}_{\vis {k_x},\vis {k_y},S\nu}(E,\vis {x},\vis {y})$ \cite{BCD09}. The wave vectors associated with the $\vis{x}$ and $\vis{y}$ scaled Jacobi coordinates are $\vis{k_x}$, $\vis{k_y}$, respectively, and $\nu$ is the projection on the $z$ axis of the total spin $S$ of the two neutrons. 

The Dirac notation in Eq.\ (\ref{E1sexact}) indicates a six-dimensional integral over the hyperspherical coordinates. The integrals over $\Omega_x$ and $\Omega_y$ can be performed analytically, but the integrals over $\alpha$ and $\rho$ require a numerical approximation.
 If we use Lagrange functions and the Gauss quadrature, the integral over $\rho$ is simply proportional to a sum over the coefficients of the expansion of the hyperradial bound and continuum wave functions.

\subsection{Four-body eikonal wave functions}
In the following we briefly describe the four-body Coulomb corrected eikonal model. For details we refer the reader to Ref. \cite{BCD09}. Let us consider a three-body projectile impinging on a target at energies much higher than the Coulomb barrier. Then, the time-independent four-body Schr\"odinger equation in scaled Jacobi coordinates is given by
\begin{eqnarray}
H_{4b}\Phi(\vis{R},\vis{x},\vis{y})=E_T\Phi(\vis{R},\vis{x},\vis{y}),
\label{se4b}
\end{eqnarray}
with 
\begin{eqnarray}
H_{4b}=H_{3b}-\frac{\hbar^2}{2\mu_{PT}}\Delta_R+V_{PT}(\vis{R},\vis{x},\vis{y}),
\end{eqnarray}
where $H_{3b}$ is the internal Hamiltonian of the three-body projectile given by Eq. (\ref{Ham3b}). The relative coordinate between the center of mass of the projectile and the center of mass of the target is $\vis{R}=(\vis{b},Z)$, with $\vis{b}$ its transverse component. The reduced mass of the projectile-target system is $\mu_{PT}$, and the total energy $E_T$ is
\begin{eqnarray}
E_T=\frac{\hbar^2}{2\mu_{PT}}k^2+E_0,
\end{eqnarray}
where $E_0$ is the ground state energy of the projectile. The initial projectile-target relative wave vector is denoted by $k$ which is defined along the $Z$ coordinate. 

The projectile-target interaction $V_{PT}$ is given by
\begin{equation}
V_{PT}=V_{cT}+V_{nT}+V_{nT},
\end{equation}
where $V_{cT}$ and $V_{nT}$ is the core-target and neutron-target potentials, respectively. 

At high energies, we can assume that the solution of the Schr\"odinger equation (\ref{se4b}) can be written as 
\begin{eqnarray}
\Phi(\vis{R},\vis{x},\vis{y})=e^{ikZ}\hat{\Phi}(\vis{R},\vis{x},\vis{y}).
\label{feik}
\end{eqnarray}
From factorization (\ref{feik}) and performing the adiabatic approximation that consists in replacing $H_{3b}$ by $E_0$ \cite{SLY03}, we get the eikonal wave function
\begin{eqnarray}
\hat{\Phi}_{\text{eik.}}(\vis{R},\vis{x},\vis{y})&=&\exp\biggl(-\frac{i}{\hbar v}\int_{-\infty}^ZdZ' \,V_{PT}(\vis{b},Z',\vis{x},\vis{y})\biggr)\nonumber\\
&&\times\Psi^{J_0M_0\pi_0}(\vis{x},\vis{y}),
\label{eikwf}
\end{eqnarray}
with $v$ the initial relative velocity between the target and the projectile. The breakup cross sections are proportional to the breakup $T$-matrix which is obtained from  the eikonal wave function (\ref{eikwf}) and is given by \cite{BCD09}
\begin{eqnarray}
T_{fi}=i\hbar v \int d^2\vis{b}\,e^{-i\vis{q}\cdot\vis{b}}S_{S\nu}(E,\vis{k}_x,\vis{k}_y,\vis{b}),
\label{tmatrix}
\end{eqnarray}
where $\vis{q}=\vis{k'}-\vis{k}$ is the transferred wave vector, $\vis{k'}$ is the final projectile-target relative wave vector and $S_{S\nu}(E,\vis{k}_x,\vis{k}_y,\vis{b})$ are the eikonal breakup amplitudes
\begin{eqnarray}
S_{S\nu}&=&\biggl( \frac{A_1+2}{A_1} \biggr)^{3/4} \nonumber \\
&&\times \bigl<\Psi^{(-)}_{\vis{k}_x,\vis{k}_y,S\nu}|e^{i\chi(\vis{b},\vis{b}_x,\vis{b}_y)}|\Psi^{J_0M_0\pi_0}\bigr >.
\label{sbup}
\end{eqnarray}
In Eq.\ (\ref{sbup}), $\chi(\vis{b},\vis{b_x},\vis{b_y})$ is the eikonal phase defined as
\begin{eqnarray}
\chi(\vis{b},\vis{b_x},\vis{b_y})=-\frac{1}{\hbar v}\int_{-\infty}^{\infty}dZ\,
V_{PT}(\vis{R},Z,\vis{x},\vis{y}),
\end{eqnarray}
where $\vis{b_x}$ and $\vis{b_y}$ are the transverse part of the scaled Jacobi coordinates. The Coulomb tidal eikonal phase leads to a logarithmic divergence of the breakup cross section \cite{SLY03}. This problem is overcome by replacing the first order of its exponential expansion by its corresponding first order of the perturbation theory (see Refs. \cite{AbS04,MBB03}). 

In practice the eikonal phase is expanded in multipoles and the excitation functions $d\sigma/dE$ can be written as a sum of different partial wave contributions \cite{BCD09}.

\section{$^{22}$C Ground state}
\label{GSproperties}
In this section, we investigate $^{22}$C properties (ground-state energy and rms radius) for various $\cn$ potentials.
These potentials provide  different scattering lengths, and different energies of a 
possible $3/2^+$ resonance in $^{21}$C \cite{De00b,OMF13,N15}.

In three-body calculations, $n-n$ and a core$-n$ potentials are needed. The $n-n$ potential is taken as
the central part of the Minnesota interaction \cite{TLT77} with a mixture parameter $u=1$. The $\cn$ potential is chosen 
as in Ref.\ \cite{KT14}
\begin{equation}
V_{^{20}\text{C}+n}(r)=-V^l_0f(r)+V_{ls}\vis {l}\cdot\vis {s}\frac{1}{r}\frac{d}{dr}f(r),
\label{core-n-pot}
\end{equation}
with $f(r)=1/\left[1+\exp(\frac{r-R_c}{a})\right]$. Parameters $a=0.65$ fm and $R_c=3.393$ fm are taken from 
Ref.\ \cite{KT14}. The depth $V_{ls}$ is fixed to 35 MeV, which is close to the values of Ref.\ \cite{HS06}. This depth is chosen to bind the $0d_{5/2}$ state at least at the neutron separation energy of $^{20}$C (2.93 MeV).
To simulate these different potentials,
we vary the depth of the $\cn$ $s$ and $d$ waves, $V_{0}^{l=0}$ and $V_{0}^{l=2}$. For all other partial waves, we adopt
$V_{0}^{l}=42$ MeV. These potentials take partly account of the Pauli principle, as they contain one forbidden state in the 
$0s_{1/2}$, $0p_{3/2}$, $0p_{1/2}$ and $0d_{5/2}$ orbitals. In the three-body calculations, the forbidden states are removed by a supersymmetric transformation \cite{Ba87}.

For the \cnn\ calculation, the ground-state wave functions (\ref{pwf}) are truncated at a maximum hypermomentum $\kmax=40$. The hyperradial wave functions are expanded in a Lagrange-Legendre basis \cite{DB10}. 
The rms matter radius is calculated with
\begin{equation}
<r^2>_{^{22}\text{C}}=\frac{20}{22}<r^2>_{^{20}\text{C}}+\frac{1}{22}<\rho^2>,
\end{equation}
where $<\rho^2>$ is the mean squared hyperradius and $\sqrt{<r^2>_{^{20}\text{C}}}=2.98\pm 0.05$ fm is the experimental rms radius of $^{20}$C \cite{OBC01}.

\begin{figure}[htb]
	\begin{center}
		\epsfig{file=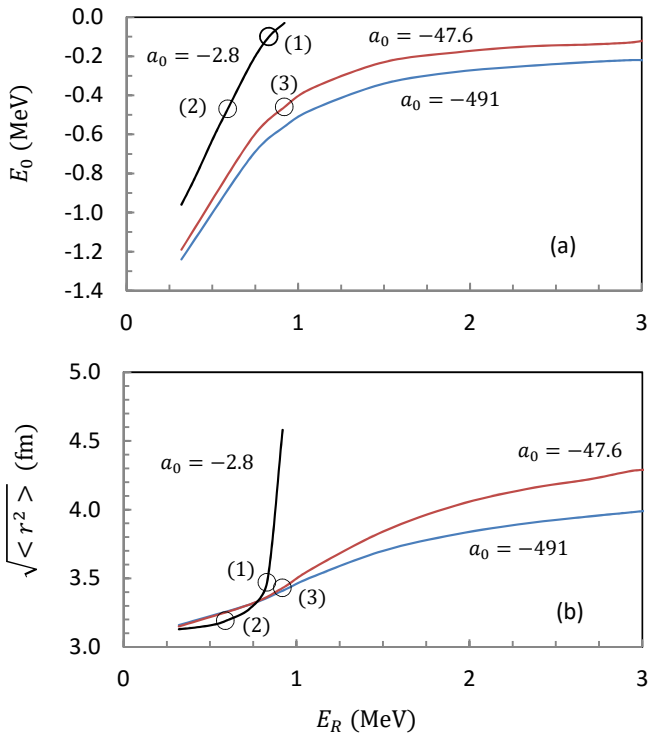,width=8.6cm}\\
		\caption{(Color online). $\carba$ energy $E_0$ (a) and  r.m.s.\ radius (b) as a function of the $0d_{3/2}$  $\carbb$ resonance energy $E_R$, and for different scattering lengths $a_0$ (in fm). The circles refer to the three potential sets.}
		\label{fig_e_r}
	\end{center}
\end{figure}

In Figure \ref{fig_e_r}, we show the dependence of the ground-state energy $E_0$ (defined from the $\cnn$ threshold) and 
of the radius as a function of the $3/2^+$ resonance energy 
$E_R$, and for various scattering lengths $a_0$ of $^{21}$C. The scattering length is directly related to $V_{0}^{l=0}$, and
is computed with the method of Ref.\ \cite{RS13}.
We consider three values: $a_0=-2.8$ fm, consistent with the data of
Ref.\ \cite{MBB13}, and two other values, larger by one and two orders of magnitude ($a_0=-47.6$ fm and $a_0=-490.7$ fm). 
These choices permit us to cover a reasonable
interval. The corresponding potential depths $V_{0}^{l=0}$ are 29.8, 33.0 and 33.5 MeV, respectively.
The amplitude $V_{0}^{l=2}$ determines the resonance energy $E_R$.

Figure \ref{fig_e_r} suggests that a very low separation energy can be obtained with a small $a_0$ only, which is consistent with the
analysis of Mosby {\sl et al.} \cite{MBB13}, who deduce $S_{2n}<70$ keV from the measured scattering
length $\vert a_0 \vert <2.8$ fm. In parallel, the large r.m.s.\ radius ($5.4 \pm 0.9$ fm) observed by
Tanaka {\sl et al.} \cite{TYS10} requires a binding energy close to zero. The value deduced by the authors ($S_{2n}=0.42\pm 0.94$ MeV)
presents a very large error bar, but large separation energies can be ruled out from the r.m.s. value.

From this first analysis, a satisfactory agreement with the available experimental data can be obtained with $V_{0}^{l=0}=29.8$ MeV and
$V_{0}^{l=2}=47.8$ MeV. These values are consistent with a large r.m.s. radius \cite{TYS10}, with a small binding energy \cite{TYS10,GMO12,MBB13},
and with the experimental $\carbb$ scattering length \cite{MBB13}. Of course, large uncertainties exist for the
binding energy, but the coherence of the different data sets favours a small value ($S_{2n}\sim 0.1$ MeV). In these conditions,
a $3/2^+$ resonance is found in $\carbb$ at $E_R=0.83$ MeV, with a width of 0.09 MeV. The existence of this $^{21}$C resonance, in parallel with
the particle stability of the ground state was suggested in Ref.\ \cite{De00b}, in the framework of a microscopic cluster model.
Preliminary experimental data \cite{En15} seem to confirm this prediction.

In addition to the $\cn$ potential mentioned before, and hereafter referred to as ``set 1", we select two other sets, which are given in Table
\ref{table_pot}. Set 2 corresponds to the same scattering length, but the $\carba$ binding energy is
larger, as suggested in Ref.\ \cite{TYS10}. With set 3, we illustrate a possibly larger scattering length.
These potentials are indicated by circles in Fig.\ \ref{fig_e_r}, and
 will be used in the next Section to compute breakup cross sections.

\begin{table}[htb]
	\caption{Parameter sets of the $\cn$ system. Energies are in MeV, and lengths in fm.
		\label{table_pot}}
	\begin{ruledtabular}
		\begin{tabular}{lccccc}
			& $V_{0}^{l=0}$ & $V_{0}^{l=2}$ & $E_0$ & $a_0$ & $E_R$  \\  
			\hline
	set 1 &	29.8 & 47.8 & $-0.10$ & $-2.8$ & 0.83 \\
	set 2 &	29.8 & 48.4 & $-0.47$ & $-2.8$ & 0.59 \\
	set 3 &	33.0 & 47.5 & $-0.46$ & $-47.6$ & 0.92 \\
		\end{tabular}
	\end{ruledtabular}
\end{table}

\begin{figure}[htb]
	\begin{center}
		\epsfig{file=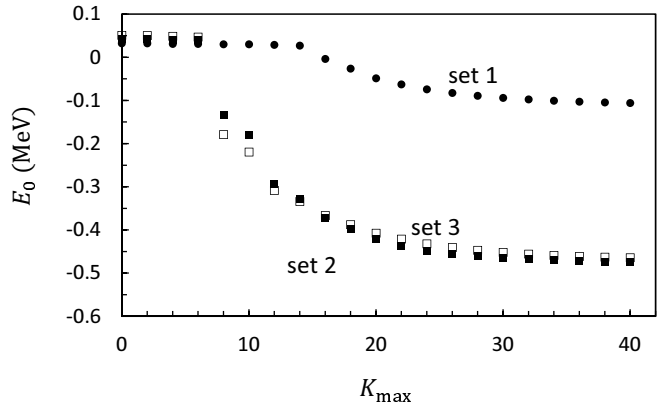,width=8.6cm}
		\caption{Convergence of the \cnn\ ground state energy $E_0$ with the maximum hypermomentum $\kmax$ for 
			the potential sets 1 (circles), 2 (filled squares) and 3 (open squares).}
		\label{kconv}
	\end{center}
\end{figure}

In Fig.\ \ref{kconv}, we illustrate the convergence of the $^{22}$C ground state energy with $\kmax$. The most weakly bound state (set 1) converges more slowly. A convergence better
that 0.01 MeV (3\%) is achieved around  $\kmax=40$. Bound-state wave functions are computed relatively fast and large
$\kmax$ values can be adopted. However, continuum three-body states are much more demanding in terms of
computer times \cite{DTB06}, and a compromise must be adopted.

Table \ref{tab_weights} shows the main contributions in the $^{22}$C ground-state wave function. These weights are defined as
\begin{equation}
C^{J\pi}_{\gamma}=\sum_K \langle \chi^{J\pi}_{\gamma K} \vert \chi^{J\pi}_{\gamma K}\rangle \backsimeq
\sum_{Ki}\vert c^{J\pi}_{\gamma K i} \vert^2,
\end{equation}
where coefficients $c^{J\pi}_{\gamma K i}$ are defined by (\ref{chivar}), and where we have used the properties of
the Lagrange functions for the matrix elements.
The sensitivity with the choice of the \textquotedblleft Y basis" (\textquotedblleft shell model like basis") and \textquotedblleft T basis" (\textquotedblleft cluster model like basis") is shown. The $\carba$ wave function
obtained with set 2 presents a different structure. In the Y basis, which emphasizes the $\carbb$ structure,
 the $d$-wave component is strongly dominant
(72.8\%), which is consistent with the low resonance energy. We therefore expect different E1 strengths and breakup cross sections with this parameter set.

\begin{table}[htb]
\begin{ruledtabular}
\setlength{\abovecaptionskip}{0pt}
\setlength{\belowcaptionskip}{10pt}
\centering\renewcommand{\arraystretch}{1.5}
\caption{Partial weights $C^{J\pi}_{\gamma}$ (in \%) of the  main components of the $\cnn$ ground state wave function potentials. The calculations are performed in the T and Y bases.}
\begin{tabular}{lrrr}
\multicolumn{4}{c}{\textbf{T basis}}\\
\hline
($S,L,l_x,l_y)$ &  set 1   & 	set 2 & set 3 \\
\hline
(0,0,0,0)   & 67.2     &  55.3 & 82.3   \\
(0,0,2,2)   & 1.7      &	2.5 & 1.0   \\
(1,1,1,1)   & 29.1      &	39.7 & 15.8    \\
(1,1,3,3)   & 1.5      &	1.9  & 0.7  \\
\hline
\multicolumn{4}{c}{\textbf{Y basis}}\\
\hline
($S,L,l_1,l_2$) &  set 1   & 	set 2 & set 3  \\
\hline
(0,0,0,0) & 35.0	&	19.7 & 59.2 \\
(0,0,1,1) & 6.3    &	4.8 & 5.2 \\
(0,0,2,2) & 26.1   &	32.1 & 17.3\\
(0,0,3,3) & 1.2    &	1.0 & 1.2 \\
(1,1,2,2) & 29.9&	40.7  & 16.0 
\label{tab_weights}
\end{tabular}
\end{ruledtabular}
\end{table}

\section{E1 strength distributions and breakup cross sections}
\label{Breakup}
\subsection{Three-body phase shifts}
We use the three-body $R$-matrix method to determine \cnn\ continuum states \cite{DTB06}. 
As the number of channels in (\ref{cde}) increases rapidly with $\kmax$, this truncation value is lower for
continuum states than for bound states. We adopt here $\kmax=30,25,20$ for the $J=0^+, 1^-, 2^+$ partial waves, respectively.
This convergence problem has been discussed in previous papers \cite{DTB06,PDB12}. 
In particular, we discussed the convergence of the E1 strength in Ref.\ \cite{PBD11}

As the electromagnetic matrix elements
involved in the breakup cross sections are sensitive to the long-range part of the wave functions, we use large Lagrange
bases with a channel radius $a\approx 90$ fm and a number of functions $N\approx 100$. 
To determine the scattering matrix $\pmb{U}^{J\pi}$ [see Eq.\ (\ref{asympchi})], the $R$-matrix is 
propagated up to 400 fm  \cite{DTB06}, owing to the long range of the potentials in hyperspherical coordinates.
Several tests have been performed to check
that the final results are insensitive to the basis choice, provided it extends to large distances with high accuracy.

\begin{figure}[htb]
	\begin{center}
		\epsfig{file=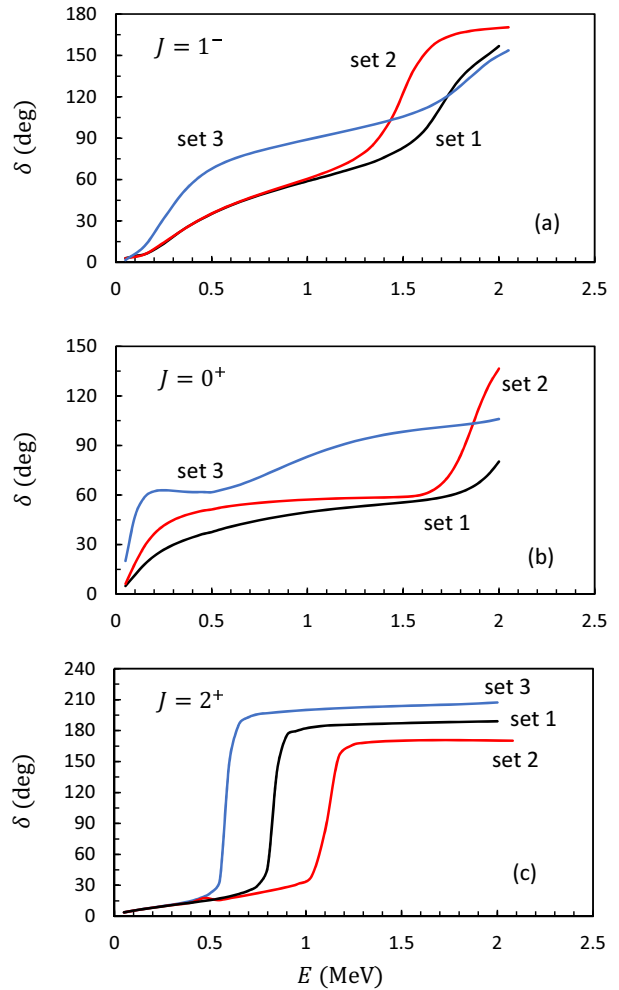,width=8.0cm}
		\caption{(Color online). Three-body $\cnn$ eigenphases for $J=1^-$ (a), $J=0^+$ (b) and $J=2^+$ (c).}
		\label{fig_delta}
	\end{center}
\end{figure}

As the breakup cross sections are expected to be dominated by the E1 contribution, the $1^-$ partial wave essentially
defines the continuum. The corresponding $J=1^-$ phase shifts are shown in Fig.\ \ref{fig_delta}(a). The scattering matrix takes large dimensions, equal to the
number of $(\gamma K)$ values (for $J=1^-\ (K_{\rm max}=25)$, the size is $260 \times 260$). Accordingly, the scattering matrix
is first diagonalized \cite{DTB06}, and the largest eigenphases are shown in Fig.\ \ref{fig_delta}.
Sets 1 and 2 present similar phase shifts, as they correspond to the same scattering length. With
set 3, however, a structure appears around 0.5 MeV.

Figures \ref{fig_delta}(b) and \ref{fig_delta}(c) show the phase shifts associated with the $0^+$ and $2^+$ partial waves,
which may affect the breakup cross sections. This is particularly true in the presence of resonances. The $2^+$ phase shift
presents a narrow resonance between 0.5 and 1.2 MeV, regardless of the potential. This result is consistent with a shell model
picture, where a $2^+$ state is predicted, based on a $(1s_{1/2})^{-1}(0d_{3/2})^1$ configuration.

\subsection{E1 strength distribution}
\begin{figure}[htb]
	\begin{center}
		\epsfig{file=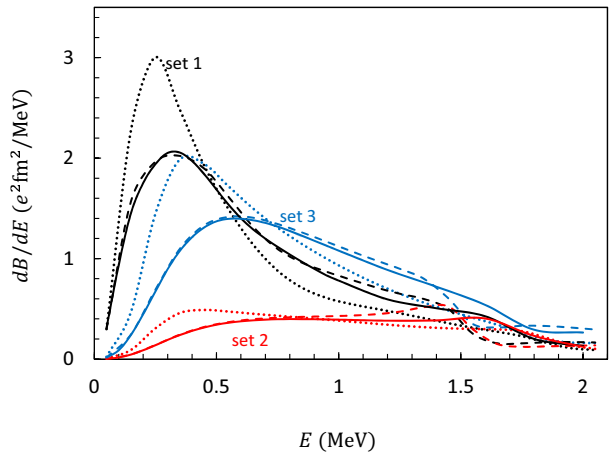,width=8.0cm}
		\caption{(Color online). Electric dipole strengths distributions of $^{22}$C. The colors correspond
			to the three potential sets for the ground state. The continuum state is defined by
			set 1 (solid lines), set 2 (dashed lines), or set 3 (dotted lines).}
		\label{fig_dbde}
	\end{center}
\end{figure}

We present in Fig.\ \ref{fig_dbde} the E1 strength distribution for the three potential sets.
Here we consider various options for the $\carba$ ground state, and for the $\cnn$ continuum
states. 

From Fig.\ \ref{fig_dbde}, it turns out that the structure of the ground state plays
the dominant role. Sets 1 and 3 provide similar E1 distributions, whereas set 2 leads to a flat curve, with a structure around 1.5 MeV. The reduction of the strength distribution with sets 2 and 3 is directly related to the
larger binding energy (see Table \ref{table_pot}). As the matrix elements in (\ref{E1sexact}) involve
an important contribution from large distances, the larger binding energy of the ground state makes the wave function
smaller at large distances. The sensitivity to the continuum state is weaker: in each case, sets 1 and 2
provide almost identical strength distributions, whereas set 3 slightly decreases the peak energy,
with an enhancement of the amplitude. 

This result confirms the conclusion of Ershov {\sl et al.} \cite{EVZ12} who use a simplified $\cn$
potential.
Clearly an experimental measurement of the strength distribution
would provide strong constraints on models, and therefore on the ground-state properties. 
\vspace*{0.5cm}

It is worth mentioning that the low-energy peak in the E1 strength is an effect of the low binding of the ground state (as we can see from Fig. \ref{fig_dbde}) and it is not a resonance effect \cite{NLV05}. 
If the peak was related with a $1^-$ resonance, it should show up at the same energy independently of the ground state.

\subsection{Breakup cross sections}

We study the $^{22}$C breakup on a $^{208}$Pb target at 240 MeV/nucleon. The $n-^{208}$Pb optical potential at 240 MeV is taken from Ref.\ \cite{KD03}. The core-target potential is the \textquotedblleft$t\rho\rho$" optical potential \cite{BHM01,HRB91} with $\alpha_{NN}=0.54$ and $\sigma_{NN}=2.75$ fm$^2$. These values are interpolated from Ref.\ \cite{BHM01}. We take the matter and charge densities of the $^{20}$C core and $^{208}$Pb target from Ref.\ \cite{CCG02}. The integrals involved in Eq.\ (\ref{tmatrix}) are solved as indicated in Ref.\ \cite{BCD09} with similar conditions. We checked that the cross sections are weakly sensitive to the potentials.

\begin{figure}[htb]
\begin{center}
\epsfig{file=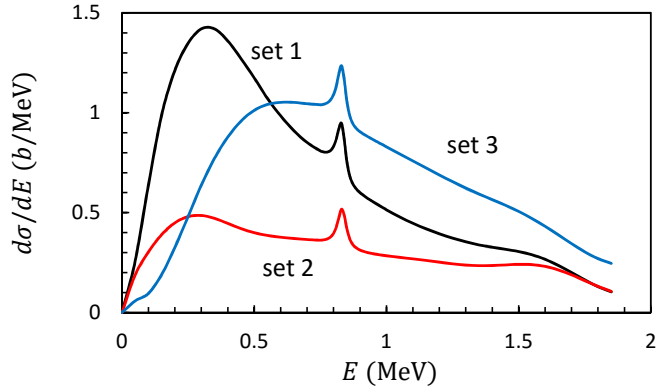,width=8.6cm}
\caption{(Color online). Total breakup cross sections of $^{22}$C on $^{208}$Pb at 240 MeV/nucleon with
	 different $\cn$ potentials (see text for details ).}
\label{fig_bu}
\end{center}
\end{figure}

\begin{figure}[htb]
	\begin{center}
		\epsfig{file=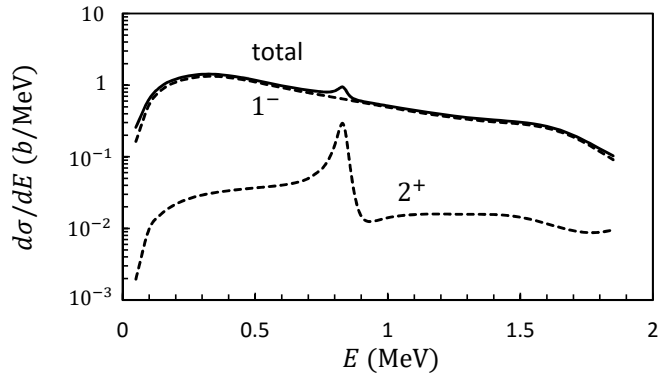,width=8.6cm}
		\caption{Total breakup cross sections of $^{22}$C on $^{208}$Pb at 240 MeV/nucleon with
			set 1 (solid lines). The contributions of the $J=1^-$ and $J=2^+$ partial waves are shown as dashed lines.}
		\label{fig_sigj}
	\end{center}
\end{figure}

For the continuum, we adopt set 1 with the $J=0^+,1^-,2^+$ partial waves, and we assess the dependence on the ground-state
properties. The cross sections are presented in Fig.\ \ref{fig_bu}. As expected, the shape of the cross section is similar to the dipole strength distribution of Fig.\ \ref{fig_dbde}.
The cross sections are fairly sensitive to the ground state properties. An interesting prediction is the presence of a
narrow peak around 0.85 MeV, and corresponding to the $2^+$ resonance (see Fig.\ \ref{fig_delta}). This is confirmed in
Fig.\ \ref{fig_sigj}, where we show the separate contributions of the $J=1^-$ and $J=2^+$ partial waves for set 1.

\section{Summary and conclusions}
\label{conclusions}
We have studied the Coulomb breakup of $^{22}$C at 240 MeV/nucleon in a four-body eikonal model \cite{BCD09}, where bound and continuum wave functions of the projectile are described in hyperspherical coordinates. This model has no free parameters, once the core$+n$ and $n+n$ potentials, necessary in the three-body model, and the core+target and $n$+target potentials, needed in the reaction framework, are fixed.

In contrast with studies on $^6$He \cite{BCD09} and $^{11}$Li \cite{PDB12} two main difficulties are faced 
in the study of the breakup of $^{22}$C: $i)$ The lack of precise experimental information of its 
ground state, $ii)$ the absence of precise knowledge of the spectroscopy of $^{21}$C. Besides, 
the very low binding energy of the  $^{22}$C ground state, $\vert E_0\vert \le 0.3$ MeV (in 
comparison with $E_0=-0.97$ MeV for $^6$He, and $E_0=-0.37$ MeV for $^{11}$Li) provides a more 
extended wave function that makes the calculations of
electromagnetic matrix elements even more time consuming.

We studied $^{22}$C properties for three $\cn$  potentials which provide plausible scattering lengths and 
energy of a possible $0d_{3/2}$ resonance \cite{De00b,En15}.  
If we consider a scattering length close to the experimental limit, $|a_0|<2.8$ fm of Ref. \cite{MBB13}, we improve the 
prediction given from a three-body zero range model \cite{MBB13}. Our calculation is more precise since it includes finite 
range two-body interactions. Therefore the limit $|a_0|<2.8$ fm implying $S_{2n}<70$ keV must be considered carefully. 

On the other hand, the position and strengths of the peaks of the dipole strength are significantly
 affected by the ground-state energy. If a extremely weakly bound state of $^{22}$C exists, it should show up from the large strength and very shifted position of the peak to low energies in the experimental breakup cross section. 

Our calculation also predicts a $2^+$ narrow resonance below 1 MeV. The energy obtained with set 1 is 0.83 MeV. Although the predicted energy
may depend on the conditions of the calculations, the existence of a $2^+$ resonance is founded from simple shell-model arguments. 
Measurements of the $^{22}$C breakup cross section would strongly help to improve theoretical models.

\section*{Acknowledgments}
We thank D. Baye for providing us with the four-body eikonal code and O. L. Ram\'{\i}rez Su\'arez for calculating the scattering lengths. E.C.P. is financed by the Fondo Nacional de Financiamiento para la Ciencia, la Tecnologia y la Innovaci\'on Francisco Jos\'e de Caldas and the Universidad Nacional de Colombia, Colombia. P.D. is Directeur de Recherches of F.R.S.-FNRS, Belgium. 
This text presents research results of the IAP programme P7/12 initiated by the Belgian-state 
Federal Services for Scientific, Technical and Cultural Affairs.

\bibliographystyle{apsrev4-1}
\bibliography{biblio2}

\end{document}